\documentclass[%
reprint,
superscriptaddress,
 amsmath,amssymb,
 aps,
 prx,
longbibliography
]{revtex4-1}

\usepackage{graphicx}
\usepackage{dcolumn}
\usepackage{bm}
\usepackage{xcolor}
\usepackage{soul}
\usepackage{acro}
\usepackage{ulem}
\usepackage{adjustbox}
\usepackage{lipsum}

\begin{document}

\title{Magnetoresistance oscillation study of the half-quantum vortex in doubly connected mesoscopic superconducting cylinders of Sr$_2$RuO$_4$}

\author{Xinxin Cai}
\altaffiliation{Present Address: School of Physics and Astronomy, University of Minnesota, MN 55455, USA.}
\affiliation{Department of Physics and Materials Research Institute, Pennsylvania State University, University Park, PA 16802, USA.}
\author{Brian M.~Zakrzewski}
\affiliation{Department of Physics and Materials Research Institute, Pennsylvania State University, University Park, PA 16802, USA.}
\author{Yiqun A.~Ying}
\affiliation{Department of Physics and Materials Research Institute, Pennsylvania State University, University Park, PA 16802, USA.}
\author{Hae-Young~Kee}
\affiliation{Department of Physics, University of Toronto, Toronto, Ontario M5S 1A7, Canada.}
\affiliation{Canadian Institute for Advanced Research, Toronto, Ontario M5S 1A7, Canada.}
\author{Manfred~Sigrist}
\affiliation{Theoretische Physik, ETH Zurich, CH-8093 Zurich, Switzerland.}
\author{J.~Elliott~Ortmann}
\altaffiliation{Present Address: Department of Physics, University of Texas at Austin, TX 78712, USA.}
\affiliation{Department of Physics, Tulane University, New Orleans, LA 70118, USA.}
\author{Weifeng~Sun}
\altaffiliation{Present Address: School of Electrical and Electronic Engineering, Harbin University of Science and Technology, Harbin, China.}
\affiliation{Department of Physics, Tulane University, New Orleans, LA 70118, USA.}
\author{Zhiqiang~Mao}
\affiliation{Department of Physics and Materials Research Institute, Pennsylvania State University, University Park, PA 16802, USA.}
\author{Ying~Liu}
\email{Corresponding author. Email: yxl15@psu.edu}
\affiliation{Department of Physics and Materials Research Institute, Pennsylvania State University, University Park, PA 16802, USA.}

\date{\today}

\begin{abstract}
The observation of the highly unusual half-quantum vortex (HQV) in a single crystalline superconductor excludes unequivocally the spin-singlet symmetry of the superconducting order parameter. HQVs were observed previously in mesoscopic samples of Sr$_2$RuO$_4$ in cantilever torque magnetometry measurements, thus providing direct evidence for spin-triplet pairing in the material. In addition, it raised important questions on HQV, including its stability and dynamics. These issues have remained largely unexplored, in particular, experimentally. We report in this paper the detection of HQVs in mesoscopic, doubly connected cylinders of single-crystalline Sr$_2$RuO$_4$ of a mesoscopic size and the examination of the effect of the in-plane magnetic field needed for the observation of the HQV by magnetoresistance (MR) oscillations measurements. Several distinct features found in our data, especially a dip and secondary peaks in the MR oscillations seen only in the presence of a sufficiently large in-plane magnetic field as well as a large measurement current, are linked to the formation of the HQV fluxoid state in the sample and crossing of an Abrikosov HQV through the sample. The conclusion is drawn from the analysis of our data using a model of thermally activated vortex crossing overcoming a free-energy barrier which is modulated by the applied magnetic flux enclosed in the cylinder as well as the measurement current. Evidence for the trapping of an HQV fluxoid state in the sample was also found. Our observation of the HQV in mesoscopic Sr$_2$RuO$_4$ provided not only additional evidence for spin-triplet superconductivity in Sr$_2$RuO$_4$ but also insights into the physics of HQV, including its spontaneous spin polarization, stability, and dynamics. Our study also revealed a possible effect of the measurement current on the magnitude of the spontaneous spin polarization associated with the HQV. 
\end{abstract}
                             
                     
\maketitle

\section{Introduction}

The presence of the long range phase coherence in a superconductor makes it possible to define the superfluid velocity, $v_s$, which can then be related to the vector potential and the gradient of the phase of the superconducting order parameter. In a doubly connected, conventional $s$-wave superconductor with a thick wall (say, larger than twice of the London penetration depth), $v_s$ is zero within the interior of the sample. Because the order parameter must be single-valued, the phase winding around the doubly connected sample must be 2n$\pi$, where n is an integer. It is straightforward to show that this leads to the formation of the full quantum vortex (FQV) whose magnetic flux is in units of $\Phi_0={h/2e}$, where $h$ is the Planck constant and $e$ is the elementary charge. If the wall thickness of the doubly connected sample is thin (thinner than twice of the London penetration depth), the superfluid velocity will be non-zero in the sample, which demands that the fluxoid (consisting the magnetic flux and a line integral of $v_s$) rather than magnetic flux itself is quantized \cite{tinkham_introduction_1996}. Experimentally, the flux and fluxoid quantizations were demonstrated long ago\cite{DeaverFairbank_Originsl_1961,DollNabauer_Originsl_1961,LittleParks_Originsl_1962}. In a type II superconductor, an Abrikosov vortex featuring a normal core possessing a magnetic flux of $\Phi_0={h/2e}$ will be found even in a singly connected bulk sample, requiring no doubly connected sample topology\cite{Abrikosov_Originsl_1957}. 

A superconducting half quantum vortex (HQV), on the other hand, is a topological object carrying a magnetic flux of $\Phi_0/2=h/4e$, as opposed to $\Phi_0={h/2e}$ of the FQV. The HQV was first predicted in superfluid $^3$He\cite{volovik_line_1976,cross_textural_1977} featuring spin-triplet pairings. The order parameter in a spin-triplet superfluid, which is often represented by the so-called d-vector, consists of the spin and the orbital parts. The direction of the d-vector represents the one onto which the projection of the Cooper pair spin is zero white its amplitude is the energy gap in the quasi-particle spectrum. It is possible for the total phase winding of 2n$\pi$ around a loop in superfluid $^3$He to be "split" equally between the spin and the orbital parts of the order parameter. The phase winding of the spin-part of the order parameter can be visualized as the the formation of a d-vector texture, which results in the reversal of the direction of the d-vector going around a loop, with the change of the phase of the spin-part of the order parameter being (2n+1)$\pi$, where $n$ is an integer\cite{kee_half-quantum_2000}. Since the phase winding in the spin part of the order parameter does not generate a mass flow, we may obtain an HQV in superfluid $^3$He from the phase winding in the orbital part of the order parameter, which will also be (2n+1)$\pi$ to ensure the total phase winding is a multiple of 2$\pi$. Alternatively, the formation of an HQV can also be understood in the so-called "equal-spin pairing" (ESP) state in superfluid $^3$He\cite{leggett_theoretical_1975}. In the ESP state, there exists a spin polarization axis with respect to which the number of the "spin-up" Cooper pairs and that of "spin-down" pairs are the same. Since Cooper pair spins are confined to the plane perpendicular to the d-vector, it is easy to see that this spin polarization axis must be perpendicular to the d-vector. When each spin species can independently carry its own vorticity, $e.g.$, the vorticity is present in only one spin species, an HQV is found\cite{vakaryuk_spin_2009}. It is easy to check that the presence of vorticity in only one spin species would require the formation of a texture of d-vector in the plane perpendicular to the equal-spin axis. 


It is clear that the same argument for the existence of the HQV can be made for a solid-state spin-triplet superconductor. Spin-triplet, $p$-wave superconductivity was proposed for Sr$_2$RuO$_4$\cite{rice_sr_1995,baskaran_why_1996}, which is the only Cu-free, layered perovskite superconductor known to date, soon after its superconductivity was discovered \cite{maeno_superconductivity_1994}. The existence of HQV in Sr$_2$RuO$_4$ was first suggested by one of the authors of the current paper and her collaborators \cite{kee_half-quantum_2000}. It was proposed that an Abrikosov FQV in a bulk Sr$_2$RuO$_4$ crystal can split into a pair of HQVs linked by a d-vector soliton, called a d-soliton. A texture of d-vector is created by the presence of the d-soliton with the direction of the d-vector reversing its direction going around its two ends, which corresponds to a phase winding of $\pi$ in the spin part of the Cooper pair wave function. However, the free energy of a pair of Abrikosov HQVs was found to depend on the ratio of the ordinary superfluid to spin fluid density, \(K = \frac{\rho_s}{\rho_{sp}} \) and lower than that of an FQV only when the ratio is within a certain range (a large K favors the stability of HQVs). As a result the HQV pair is expected to be stable only within a small range of temperature below T$_c$. It was pointed out\cite{chung_stability_2007,kee_half_2007} that the free energy of a pair of HQVs is higher than a single FQV primarily because the unscreened charge-neutral spin fluid circulating the HQV diverges logarithmically with the sample size. This energy can be reduced by geometric constraints such as those found in a slab. However, the interaction among HQVs in a confined sample geometry is complicated and not understood.  

The formation of an HQV fluxoid state in a doubly connected, spin-triplet superconducting cylinder was also considered\cite{chung_stability_2007, vakaryuk_spin_2009, kee_releasing_2013}, which avoids the complication of the normal core in an Abrikosov HQV. Similar to Abrikosov HQV, an HQV fluxoid state also depends on $K$ as well as material and sample geometry parameters. In particular, it was found that the presence of a d-soliton in the doubly connected sample can favor the stability of the HQV fluxoid state\cite{kee_releasing_2013} (see below). Furthermore, it was found that the presence of spin currents must be accompanied by a spontaneous spin polarization (SSP). In a doubly connected sample, the velocity mismatch between the two condensate components in the ESP state will lead to a difference in the vorticity of the two spin species and an SSP in the direction of the spin polarization axis, which is perpendicular to the direction of the d-vector\cite{vakaryuk_spin_2009}. 

Experimentally, the existence of the HQV fluxoid state was detected in doubly connected, single crystalline cylinders of Sr$_2$RuO$_4$ with a mesoscopic size by cantilever torque magnetometry measurements \cite{jang_observation_2011}. Ramping up the $c$-axis magnetic field was found to induce jumps in the $c$-axis magnetization of the sample with a regular height corresponding to that of an FQV. However, in the presence of an in-plane magnetic field, and only in the presence of it, two half-height jumps instead of a single one were observed near (applied) half-flux quanta\cite{jang_observation_2011}. Such two-jump features in the magnetization can be interpreted $only$ by the emergence of HQVs in the presence of an in-plane field and the role of the necessary in-plane magnetic field must be to stabilize the HQV. It is interesting to note that this method of detecting the HQV followed the original torque measurement of the FQV in conventional superconductors\cite{DollNabauer_Originsl_1961}, featuring however a vastly improved detection sensitivity.
The discovery of HQVs in mesoscopic Sr$_2$RuO$_4$ raised important questions on the stability of HQVs. The absence of the half-step feature in the $c$-axis magnetization when an in-plane magnetic field is not applied or too small\cite{jang_observation_2011} suggests that the mesoscopic sample size alone is not sufficient to stabilize the HQV. Evidently, the spontaneous spin polarization predicted to accompany the HQV\cite{vakaryuk_spin_2009} can be used to lower the free energy of the HQV through its coupling to the in-plane field, $\Delta F=-\mu_x|\mu_0 H_{||ab}|$, where $\mu_x$ is the in-plane spontaneous spin magnetic moment and $\mu_0$ is vacuum permeability. Furthermore, as the HQV is stabilized by an in-plane magnetic field via Zeeman coupling, and furthermore, the direction of the in-plane field can be along any in-plane direction without affecting the stability of the HQV fluxoid state, the d-vector will most likely be along the $c$ axis even though it can also be along the in-plane direction perpendicular to the applied in-plane field. 



Following the initial discovery of flux quantization by flux and torque magnetometry measurements in conventional $s$-wave superconductors \cite{DeaverFairbank_Originsl_1961,DollNabauer_Originsl_1961}, measurements on the $T_c$ or magnetoresistance (MR) oscillations  were carried out by Little and Parks (LP)\cite{LittleParks_Originsl_1962}, deepening the understanding of the flux or fluxoid quantization. Similarly, MR oscillations measurements on mesoscopic cylinders of Sr$_2$RuO$_4$ will provide further understanding of HQV physics. A proposal for MR detection of the HQV in a perforated thin film of Sr$_2$RuO$_4$ featuring closely placed, periodic holes modeled by a heavily damped Josephson junction array (JJA) was put forward several years ago\cite{vakaryuk_effect_2011}.
Under an applied magnetic field, the measurement current induced phase slips, which can be viewed as thermally excited crossings of vortices, result in a field-dependent voltage, or MR, which was calculated originally by Ambegaokar and Halperin (AH)\cite{AH_model_1969}, yielding 
\begin{equation}
\label{eq:AH}
R/R_N=I_0^{-2}\left(\Delta E/2k_B T\right),
\end{equation}
where $R_N$ is the sample resistance in the normal state, $\Delta E$ is the free-energy barrier for the vortex crossing, and $I_0(x)$ is the zero-order modified Bessel function of the first kind \cite{vakaryuk_effect_2011}. Note that phase slips over a Josephson junction or nanowire of a spin-singlet superconductor and those of a corresponding sample of a spin-triplet superconductor may be different in detail. 
Nevertheless the spin-singlet result shown in Eq.~\ref{eq:AH} will still be useful for obtaining a qualitative understanding of the MR data. 

The rest of the paper will be organized as follows. In sect. II, we will present the experimental methods, including details on the sample preparation and measurements; in sect. III, we will present our experimental results, including distinct features found in the MR oscillations of doubly connected, single crystalline cylinders of Sr$_2$RuO$_4$ with a size similar to those used in the original torque magnetometry measurements in the presence of a sufficiently large in-plane magnetic field. We will present the analysis of the data and suggest that these features are consequences of the crossing of HQVs, and sometimes, being trapped in it; in sect. IV, we will discuss the implications of our results before we conclude in the final section. 


\section{Experimental Methods}


\begin{table*}
\caption{\textbf{Summary of parameters of samples used in the current experiment.} Parameters characterizing the size of the doubly connected cylinder of Sr$_2$RuO$_4$ were estimated from scanning electron microscope (SEM) images. Other parameters were obtained from the electrical transport measurements. These parameters include mid-point radius $r$, the wall thicknesses on the top and the bottom of the cylinder, $w_1$ and $w_2$, respectively, the cylinder height $h$, the superconducting transition temperatures for zero resistance, $T_{c,R=0}$, and for the onset, $T_{c,onset}$, and values of the critical current density at fixed temperatures. 
}
\begin{ruledtabular}
\begin{tabular}{cccccccr}
& & \multicolumn{2}{c}{$w$~(nm)} & & \multicolumn{2}{c}{$T_c$~(K)} & \\ \cline{3-4} \cline{6-7}
Sample & $r_m$~(nm) & Top & Bottom & $h$~(nm) & $R=0$ & Onset & $J_c$~($10^3$A/cm$^2$)  \\
\hline
HL & 601 & 191 & 334  & 780 & 0.99 & $\geq$1.63 & 8.6 at $T$= 0.5~K  \\
BL & 468 (full: 534) & 133 (full: 268) & 283 (full: 414) & 644 & 1.18 & $\geq$1.4 & 24 at $T$= 0.3~K  \\
E & 565 & 190 & 298 & 482 & 0.71 & $\geq$1.35 & 92 at $T$= 0.3~K \\
B & 597 & 183 & 322 & 540 & 1.04 & 2.1 & 75 at $T$= 0.55~K 
\end{tabular}
\end{ruledtabular}
\label{table1}
\end{table*}

\begin{figure}
\includegraphics{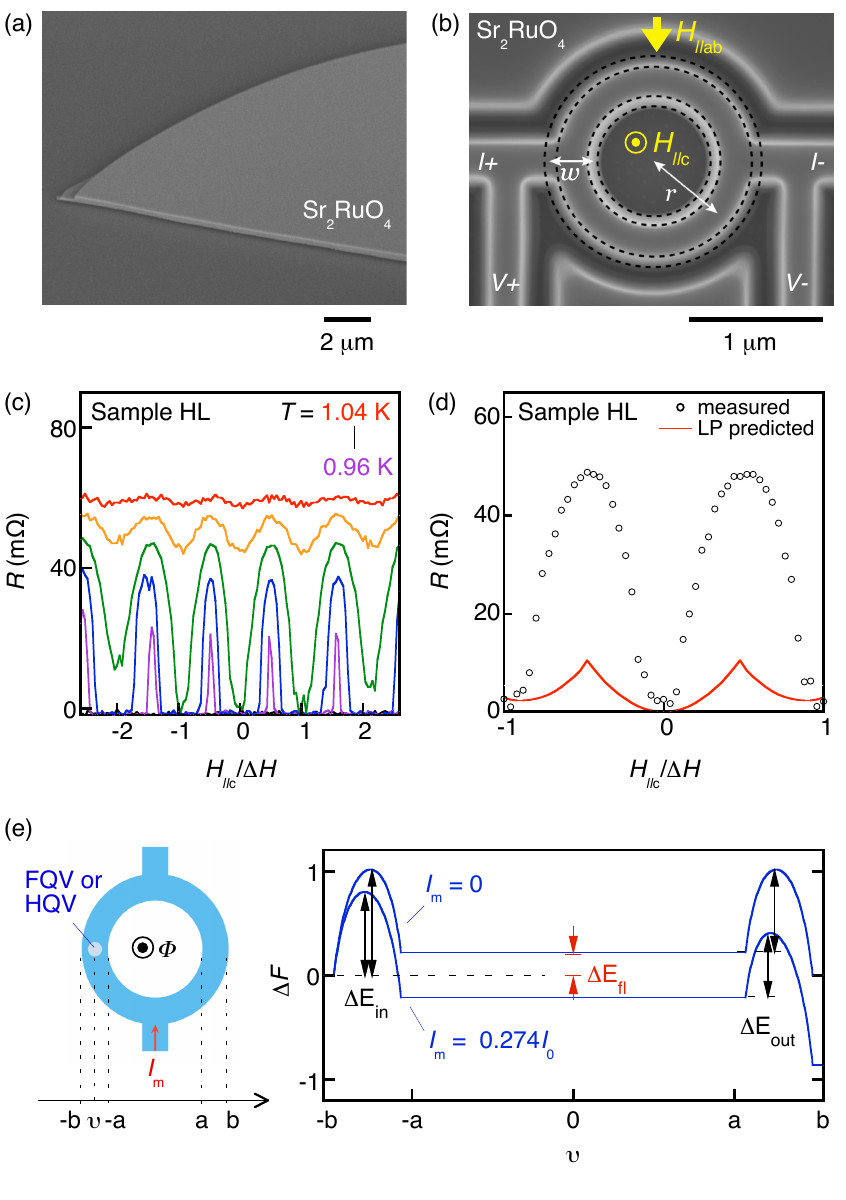}
\caption{\label{Fig1} \textbf{Mesoscopic samples of Sr$_2$RuO$_4$ and magnetoresistance (MR) oscillations induced by vortex crossing.} (a) A scanning electron microscope (SEM) image of a thin single crystal of Sr$_2$RuO$_4$ obtained by mechanical exfoliation. (b) SEM image of a doubly connected cylinder with four measurement leads of Sr$_2$RuO$_4$ cut from a thin crystal using focused ion beam (FIB) (a top view). Directions of the $c$-axis and in-plane magnetic fields, $H_{||c}$ and $H_{||ab}$, respectively, are indicated. (c) MR oscillations, $R~vs.~H_{||c}$ at various temperatures close to $T_c$ for Sample HL, where $R$ (= [$V(I=I_m)-V(I=0)]/I_m$) is the sample resistance, $I_m$ is the dc measurement current, and $\Delta H_{||c}$ is the oscillation period. Here $H_{||ab}$ = 0, $I_m = 10~\mu$A, and $\Delta H_{||c} = 15.4~$Oe. From top to bottom, $T$ = 1.04, 1.02, 1, 0.98, 0.96 K. (d) MR oscillations measured at $T = 1~$K (open circles) and calculated from Little-Parks (LP) effect (solid lines) using the sample dimensions and $R$($T$) measured at $I_m = 10~\mu$A (see main text). 
(e) Left: Schematics of a doubly connected cylindrical sample. An Abrikosov FQV or HQV (filled circles) is shown to enter the cylinder wall from the left. The fluxoid states, which are indicated by the phase winding numbers for orbital and spin parts of the order parameter, $(n_s, n_{sp})$, will transition to $(n_s',n_{sp}')$. The fluxoid state of the cylinder will return to the original one when an HQV or FQV of the sign opposite to one entering the cylinder exits the sample. Right: Examples of the free energy barrier for vortex crossing, which were calculated by placing an FQV in a ring of $s$-wave superconductor shown on the left. The applied magnetic flux was set at $\Phi$ = $\Phi_0$/4. Values of $I_m$ are indicated where $I_0$ is the maximal circulating current seen at $\Phi$ = $\Phi_0$/2 (main text). 
}
\end{figure}

High-quality bulk single crystals of Sr$_2$RuO$_4$ were synthesized by the floating-zone method. To fabricate mesoscopic samples for the MR oscillations measurement, we utilized a thin crystal plate of Sr$_2$RuO$_4$ obtained by mechanical exfoliation from a bulk single crystal along the $ab$ plane, which was then placed on a Si/SiO$_2$ substrate, as reported previously in \cite{cai_unconventional_2013}. A typical thin crystal, shown in Fig.~\ref{Fig1}(a). The typical lateral dimension of our thin crystals is 10-50~$\mu$m and the typical thickness is 300-800~nm. 

Our device fabrication began by the preparation of the electrical contacts to the crystal using photolithography which defined the pattern of the electrical contacts. Electron beam evaporation was used to deposit a layer of 10~nm thick Ti followed by a 200~nm thick Au after the surface of the crystal was cleaned by ion milling. Ti/Au contacts were evaporated from an angle (about $45^{\circ}$ to the normal of the substrate) to ensure continuous coating of the metals on the side surfaces of the crystal . A 200~nm thick SiO$_2$ was deposited on top of the crystal plates as a protective layer before the doubly connected cylinder with four electrical leads was cut by a focused ion beam (FIB) (using 30~keV Ga ions and a beam current of 50~pA). A typical structure of the cylindrical samples is shown in Fig.~\ref{Fig1}(b). The central axis of the cylinders is along the $c$-axis of the thin single crystal of Sr$_2$RuO$_4$. 

The original magnetometry detection of HQVs were carried out on doubly connected, single-crystal samples of Sr$_2$RuO$_4$ with a mesoscopic size. However, the samples used in that experiment featured a large and uneven wall thickness. While the inner edge of the sample, the center hole, was cut by FIB, the outer edge was not cut at all \cite{jang_observation_2011}, probably due to concerns on potentially destroying superconductivity from the cutting by a high-energy FIB. For samples used in the present work, both the inner and outer edges of the cylinder were cut by FIB to form the cylinder that is appropriate for the detection of HQVs by MR oscillations measurements. The use of the 200~nm thick protective layer of SiO$_2$ was important to prevent conducting material cut off from the crystal from being redeposited on the top of the cylinder which may affect our transport measurements. The use of the 200 nm thick protective layer of SiO$_2$ was important to minimize the damage of the cylinder from the Ga implantation by the high-energy ion beam.

Previous transmission electron microscope (TEM) studies of FIB-cut Sr$_2$RuO$_4$ cylinders\cite{cai_unconventional_2013} revealed that the cylinder wall features a gradually increasing wall thickness from top to bottom. Because the upper part of the crystal was exposed to the FIB longer than the lower part during the cut, a gradually increasing wall thickness is expected. In addition, TEM images showed that the inner and the outer walls of the cylinder both feature an FIB damaged surface layer of about 20~nm thick. This FIB damaged surface layer is not likely to be superconducting given the sensitivity of superconductivity in Sr$_2$RuO$_4$ to disorder. The dimensions of the cylindrical samples listed in Table~1, which were estimated based on scanning electron microscope (SEM) images, were obtained by subtracting the 20 nm surface layer from both the inner and the outer surfaces of the cylinder wall. 

The cylindrical samples were measured in a dilution refrigerator with a base temperature of 20~mK. Four-terminal resistance, obtained via a standard dc technique using a measurement current of $I_m$, was calculated by $R = [V(I=I_m)-V(I_m=0)]/I_m$. The current was supplied by a Keithley 220. The voltage across the sample was measured using a Keithley 2182. The voltage offset at $I_m$=0, which may come from sources such as the contacts between two different materials, Au/Ti and Sr$_2$RuO$_4$, as well as background signals in the measurement circuitry, is subtracted to obtain the true response of the sample to the measurement current flowing through the sample in the specified direction.  

An in-plane magnetic field, $H_{||ab}$, was applied using a superconducting solenoid magnet. A $c$-axis magnetic field, $H_{||c}$, was applied using a homemade superconducting Helmholtz coil installed inside the large superconducting solenoid. The current in the Helmholtz coil was provided by a Keithely 2400 Sourcemeter, generating a magnetic field about 30.8~Oe per 0.1~A of magnet current. The sample planes were aligned manually with respect to the magnetic field directions as precisely as possible. A typical misalignment of 1-2 degrees, possibly larger for some samples, is expected. 


The applied magnetic flux ${\it\Phi}$ enclosed in the doubly connected cylinder is determined by the $c$-axis magnetic field, $H_{||c}$. The $H_{||c}$ period for the primary MR oscillations, corresponding to a full flux quantum, $\Phi_0$, can be calculated from the cylinder dimensions using the relation $\Delta H= \Phi_0/[\pi r^2(1+(\frac{w}{2r})^2)]$, where $r$ is the mid-point radius. A correction term of order $(\frac{w}{2r})^2$ will also be included using an average wall thickness $w$ \cite{Groff_fluxoid_1968}. Experimentally, the $H_{||c}$ period for the MR oscillations was determined from the average of the peak-to-peak and valley-to-valley values of the first period at the highest temperature for which the MR oscillations data is available for the sample because MR oscillations at these conditions were seen to be most smooth.  

For samples of Sr$_2$RuO$_4$ with a mesoscopic size the values of the onset $T_c$ were found to be higher than the bulk phase, 1.5~K. The enhanced $T_c$ value could be related to the presence of Ru inclusions, uniaxial strains, and dislocations \cite{ying_enhanced_2013}, or, as to be discussed below, the presence of multiple branching points in the sample. However, in the current study, no Ru inclusions were seen in our samples under either the optical or the SEM images. The absence of the Ru inclusion is consistent with the fact that the thin crystals of Sr$_2$RuO$_4$ could be obtained by mechanical exfoliation only in crystals with a stochiometry that is either optimal or slightly Ru deficient. It is known that Ru-rich crystals of Sr$_2$RuO$_4$ cannot be cleaved.  

\section{Results}

\begin{figure*}[t]
\includegraphics[width=\textwidth]{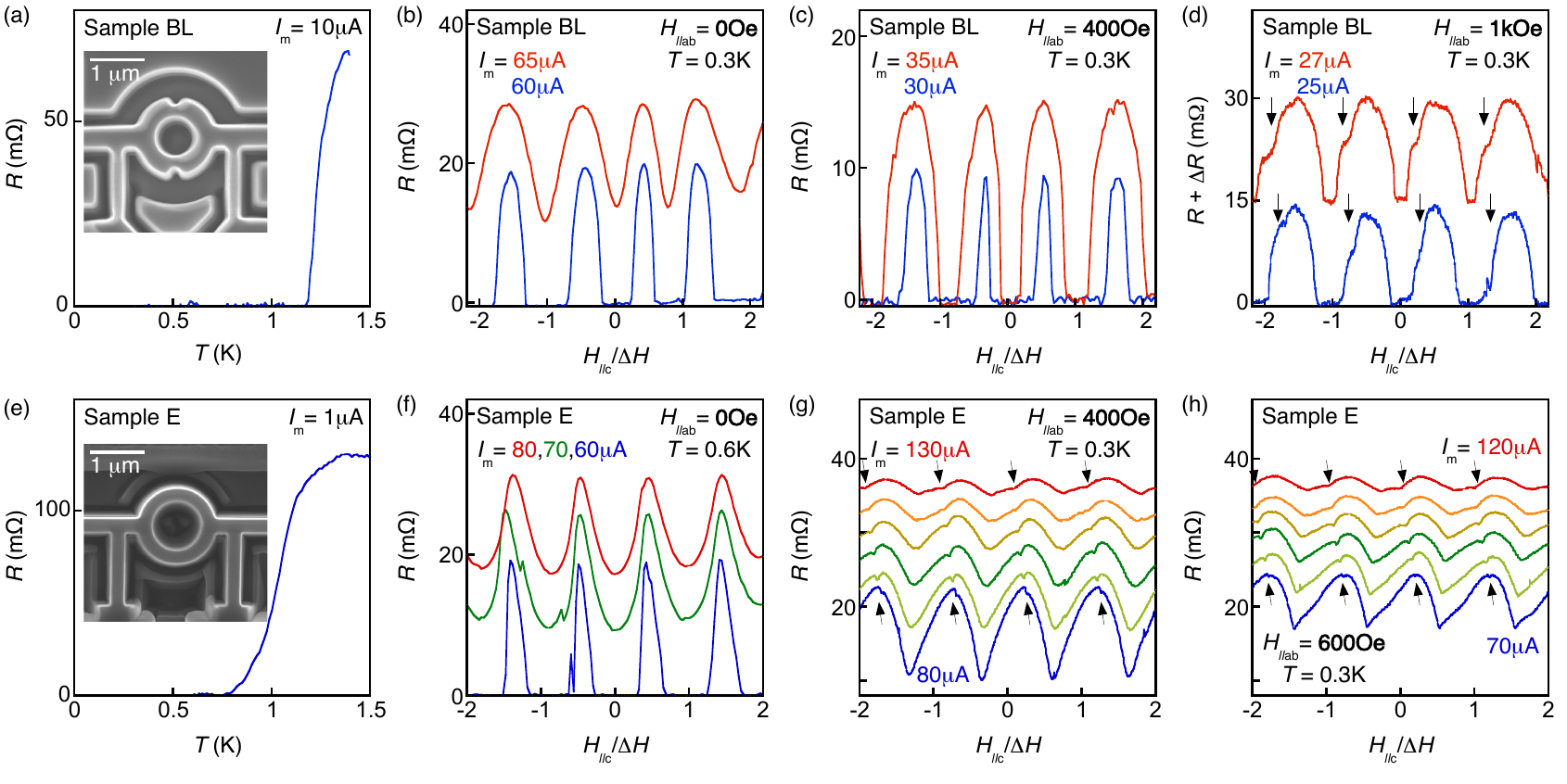}
\caption{\label{Fig2} \textbf{Dip feature in MR oscillations}. (a) Sample resistance ($R$) $vs$.~temperature ($T$) for Sample~BL, measured at $I_m = 10~\mu$A and zero magnetic field. Inset: SEM image of the sample featuring two geometric constraints. (b-d) $R$~$vs$.~$H_{||c}$ measured at fixed in-plane magnetic fields, $H_{||ab} = 0$ (b), 400 (c), and 1000~Oe (d) with values of the measurement current, $I_m$, indicated. The top curve in (d) was shifted vertically by 15~m$\Omega$ for clarity. The oscillation period is $\Delta H_{||c} = 19.7~$Oe for Sample~BL. (e) $R~vs.~T$ of Sample~E at zero magnetic field measured at $I_m = 1~\mu$A. Inset: SEM image of the sample. (f-h) $R~vs.~H_{||c}$ were measured at fixed in-plane fields of $H_{||ab} = 0$ (f), 400 (g), and 600~Oe (h) with values of $I_m$ indicated in (f). For (g) and (h), the values of $I_m$ are, from top to bottom, 130,120, 110, 100, 90, and 80 $\mu$A in (g) and 120, 110, 100, 90, 80, 70 $\mu$A in (h). $\Delta H_{||c}=18.8~$Oe for Sample~E. Except (d), no shift was made in all other plots. The dip features are indicated by arrows. 
}
\end{figure*}

\subsection{Magnetoresistance oscillations}

We measured the sample resistance as a function of $H_{||c}$ in zero $H_{||ab}$ and observed pronounced MR oscillations as shown in Fig.~\ref{Fig1}(c) for Sample HL. The period of the MR oscillation, $\Delta H_{||c}$, is in good agreement with that estimated from $\Phi_0$ using the sample dimensions listed in Table 1.
The amplitude of the MR oscillations was found to be much larger than that expected from the conventional LP effect\cite{LittleParks_Originsl_1962,Groff_fluxoid_1968}, as seen previously\cite{cai_unconventional_2013}. The $T_c$ oscillations in the LP effect lead to MR oscillations with an amplitude of $\Delta R(T) = \Delta T_c (dR/dT)$, where $dR/dT$ is the slope of the $R$~$vs$.~$T$ curve, in the transition region and $\Delta T_c$ is given by the standard formula found in literature \cite{Groff_fluxoid_1968,cai_unconventional_2013}. 
In Fig.~\ref{Fig1}(d), we plot the amplitude of the observed MR oscillations and that estimated from the LP effect for comparison. In addition, the MR oscillations were found to  persist down to a very low temperature, also different from the LP resistance oscillations which are typically seen only near $T_c$ (in zero magnetic field). The large amplitude of MR oscillations and their persistence down to a low temperature suggest that the observed MR oscillations are induced by vortex crossing\cite{cai_unconventional_2013}, observed previously in samples of Type II $s$-wave superconductors as well \cite{sochnikov_large_2010,berdiyorov_large_2012,mills_vortex_2015}. 

According to Josephson relations, the vortex crossing leads to phase slips across the sample, which in turn induce a finite voltage, $V$, in the direction perpendicular to the direction of the vortex crossing\cite{tinkham_introduction_1996}. The magnitude of $V$ is proportional to the rate of phase slips (or vortex crossing), as shown for a heavily doped Josephson junction\cite{AH_model_1969}. Even though $V$ usually depends on $I_m$ nonlinearly, $R$ = $V$/$I_m$ is still defined as the sample resistance at the specified $I_m$ value. Following the established convention, the magnetic field dependence of $R$ is referred to as MR. 

\subsection{Current driven vortex crossings}

For a vortex to cross a mesoscopic sample used in the current experiment, the magnetic flux needs to enter the sample and form a $c$-axis Abrikosov FQV or HQV, which is possible when the thickness ($w$) of the cylinder wall is larger than twice the superconducting coherence length, or the size of the normal core of the Abrikosov vortex ($\approx2\xi_{ab}(0)\approx132~$nm). This condition is satisfied in all samples used in the present study. Driven by the measurement current, $I_m$, the Abrikosov FQV or HQV formed in the cylinder wall can move into the interior of the doubly connected cylinder as shown in Fig.~\ref{Fig1}(e) (Left panel). 

The fluxoid state of the cylinder, denoted by $(n_s,n_{sp})$, where $n_s$ and $n_{sp}$ are the phase winding numbers for the orbital and the spin parts of the order parameter, respectively, is subject to change during the vortex crossing. Note that both $n_s$ and $n_{sp}$ can be an integer or half an integer, but their sum needs to be an integer in order to ensure that the total phase winding of the order parameter is a multiple of 2$\pi$. The crossing of the Abrikosov HQV (or FQV) 
will result in a fluxoid state transition from $(n_s,n_{sp})$ to $(n_s',n_{sp}')$, with the change in $(n_s)$ half an integer for the crossing of an HQV and that of an FQV an integer. Next, the flux enclosed in the interior of the cylinder will enter the cylinder wall opposite to the entry side, which will form an Abrikosov FQV or HQV again, be driven across the wall, and exit. The fluxoid state will change from $(n_s',n_{sp}')$ back to $(n_s,n_{sp})$. 

This free-energy barrier for the HQV or FQV to cross a cylinder of spin-triplet superconductor by thermal activation has not been treated theoretically. However, the corresponding barrier for FQVs in a doubly connected type II $s$-wave superconductor was calculated in the London limit\cite{kogan_properties_2004} in which the normal core of the FQV is ignored. Below we will base our discussion of the "potential barrier" for an HQV or FQV crossing in a spin-triplet superconductor on the calculations for the FQV crossing in a spin-singlet sample in this limit. The qualitative features for both processes should be similar. 

Generally speaking, the "potential barrier" is a function of vortex position in the cylinder wall, $v$, magnetic flux enclosed in the cylinder, $\Phi$, and the measurement current, $I_m$. At fixed $\Phi$ and $I_m$, the barrier consists of three segments as shown in Fig.~\ref{Fig1}(e): Segment~1, the free energy of an FQV or HQV in the "entry" side of the cylinder wall into which the applied magnetic flux penetrates. This part of the vortex position dependent free energy features a maximal value, $\Delta E_{in}$, which can be consider the "potential barrier" for the FQV or HQV to cross the wall; Segment~2, the free energy of the fluxoid state becomes $(n_s',n_{sp}')$ after the FQV or HQV enters the interior of the cylinder, which is an "excited state" as opposed to the initial state of $(n_s,n_{sp})$. The free energy difference between them is denoted by \(\Delta E_{fl} = F(n_s',n_{sp}') - F(n_s,n_{sp}) \); and Segment~3, the free energy of the vortex in the opposite side of the cylinder wall into which the magnetic flux enters to form an FQV or HQV. This part of the free energy also features a maximal value, $\Delta E_{out}$, relative to the free energy of the fluxoid state of $(n_s',n_{sp}')$, which is the "potential barrier" for the vortex crossing through this "exit" side of the cylinder wall. The fluxoid state of the cylinder will return to $(n_s,n_{sp})$ at the end of this four-step process. 

Values of $\Delta E_{in}$ and $\Delta E_{out}$ for the FQV or HQV are both relative to that before they form in the wall. In the thin-wall limit, the vortex formation in the cylinder wall is not allowed. As a result, the free energy barrier for the vortex crossing will be determined solely by $\Delta E_{fl}$.  

As $\Phi$ is varied, the circulating supercurrent in the cylinder, $I_s$, is varied as a function of $\Phi$ periodically because of the requirement of gauge invariant phase winding \cite{tinkham_introduction_1996}. This periodically varying $I_s$ will lead to the corresponding variation of the kinetic (or field gradient) part of the free energy as a function of $\Phi$\cite{kogan_properties_2004}, which will lead to the periodic variation of the "potential barrier" for the vortex crossing as to be discuss in more detail below. In the presence of $I_m$, a work will be done on the crossing vortex by the Lorentz force inserted by $I_m$ in the direction perpendicular to it. The effect of $I_m$ can then be incorporated in the free energy of the system phenomenologically by adding this work to the free energy, which will then “tilt” the "potential barrier" as shown in Fig.~\ref{Fig1}(e). It is important to note that at any fixed $I_m$, the "potential barrier" varies periodically as a function of $\Phi$. 

Note that vortex crossings in both directions perpendicular to $I_m$ are allowed in principle, with different rates, which will induce voltages in the sample with opposite signs. Experimentally measured $V$, and therefore the sample resistance, $R$=$V$/$I_m$, will be determined by the net rate of vortex crossing. Importantly, $R$ will be a periodic function of $\Phi$, leading to vortex crossing induced MR oscillations. 

\subsection{In-plane field induced dip feature in MR oscillations}


It was shown that the kinetic energy of the spin current of an HQV fluxoid state in a cylinder featuring constrictions where small parts of the wall are thinner than the rest of it will be determined by the smallest rather than the average wall thickness\cite{jang_observation_2011}. Taking advantage of this effect, we prepared a cylinder, Sample BL, featuring two intentionally made constrictions with the rest of the cylinder wall relatively thick [Inset of Fig.~\ref{Fig2}(a)]. Such sample geometry should minimize the damage of the sample by the high-energy FIB. At the same time, the chance for the HQV to form in the sample should be maximized. The sample resistance $R$ was measured as a function of $H_{||c}$ at constant in-plane fields using different values of $I_m$. Pronounced MR oscillations with a period of $\Phi_0$ were observed [Fig.~\ref{Fig2}(b-d)]. A small but clear dip feature was seen in MR oscillations at $H_{||ab} = 1000$~Oe [indicated by the arrows on the side of the MR peaks in Fig.~\ref{Fig2}(d)], which is not seen at $H_{||ab} = 0$ [Fig.~\ref{Fig2}(b)] or 400 Oe [Fig.~\ref{Fig2}(c)].



MR oscillations were measured in another sample, Sample E, which possesses a uniform wall thickness comparable with that at the constrictions in Sample BL [Fig.~\ref{Fig2}(e)]. At $H_{||ab}=0$, only smooth MR oscillations with a period of $\Phi_0$ were observed [Fig.~\ref{Fig2}(f)]. At $H_{||ab}=400$ and 600~Oe, a dip feature is seen clearly on each resistance peak of the MR oscillations, as indicated by arrows in Figs.~\ref{Fig2}(g) and (h).

\begin{figure}[t]
\includegraphics{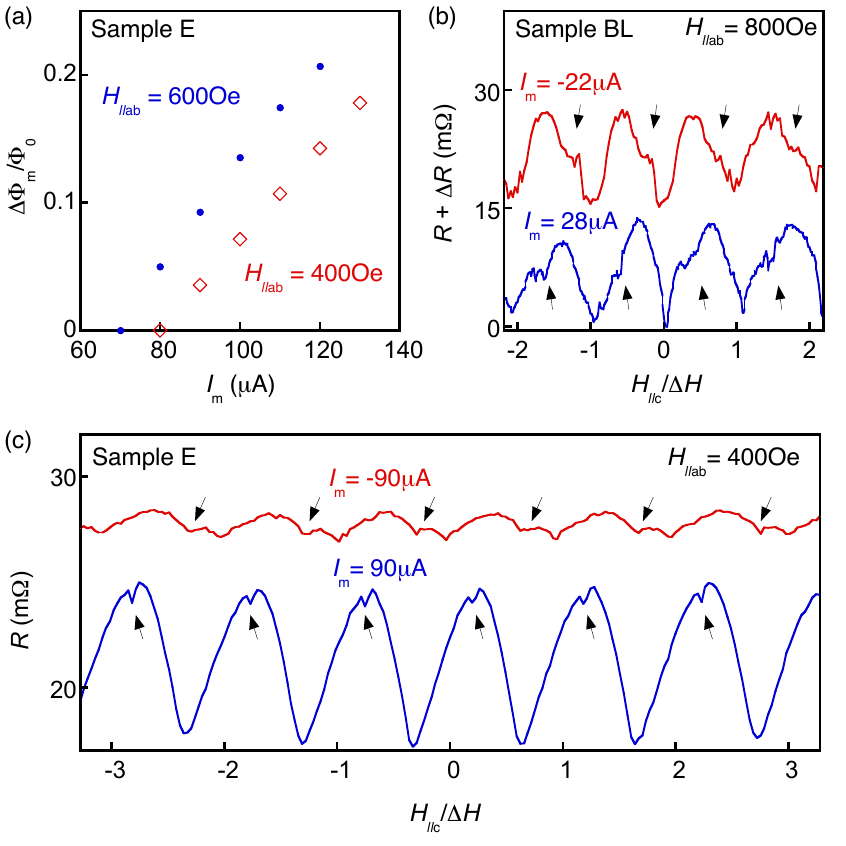}
\caption{\label{Fig3} \textbf{Effect of the measurement current on the dip feature in MR oscillations}. (a) Position of the dip feature ($\Phi = \Phi_m$) $vs$.~measurement current ($I_m$) for Sample~E obtained at in-plane fields of $H_{||ab} = 400$ and 600~Oe. The shift is measured from the maximum of the main MR peak ($R$=$R_{max}$). (b-c) $R$~$vs$.~$H_{||c}$ obtained at a fixed $H_{||ab}$ and $I_m$ as indicated for Samples BL and E. The top curve in (b) is shifted vertically by 15~m$\Omega$ for clarity. Reserving the direction of $I_m$ leads to a shift of the dip feature in the MR oscillations from one side of the main MR peak to another (indicated by the arrows) as well as a significant change in the amplitude of MR oscillations for Sample E.
}
\end{figure}

The MR oscillations data shown in Figs.~\ref{Fig2}(g) and (h) revealed several interesting trends. First, at a fixed $I_m$, the dip feature is seen to locate on the same side of the MR peak independent of the sign of $H_{||c}$ or $\Phi$. When the direction of $I_m$ is reversed, the dip feature switches from one side of the main MR peak to the other, a trend seen in both Samples BL and E [Figs.~\ref{Fig3}(b) and (c)]. Second, the position at which the dip feature was found, referred to as $\Phi = \Phi_m$, is very close to the center of the main peak at some small values of $I_m$, moving away from the center of the main MR peak as $I_m$ was increased [Figs.~\ref{Fig3}(a)]. Finally, the amplitude of MR oscillations was found to depend not only on the magnitude but the direction of $I_m$ as well for Sample E, suggesting that $I_m$ play a crucial role in determining the rate of vortex crossings.

Several other observations should also be noted. For Sample E, at $H_{||ab}$=0, minimal values of MR appear to locate near applied integer flux quanta, \(\Phi = m\Phi_0 \), where $m$ is zero or an integer [Fig.~\ref{Fig2}(g)], as expected. However, for $H_{||ab}$ = 400 and 600~Oe, MR minima were seen to be away from full applied flux quanta, \(\Phi=m\Phi_0 \), which may be due to a combination of the misalignment of $H_{||ab}$ and possibly intrinsic reasons as well. For Sample HL, which features a wall thickness similar to Sample E but a height larger than both Samples E and BL, the dip feature was either not stable or missing even under a large $H_{||ab}$, as shown in more detail in SM\cite{SM}. This may be due to its large sample height (Table~1). The crossing of $c$-axis HQVs or FQVs will tend to be affected more strongly by sample details. In-plane vortices will also be formed more favorably in this sample than in others. These factors will tend to affect the dip feature, as to be discussed in more detail in SM\cite{SM}.





\begin{figure*}[t]
\includegraphics{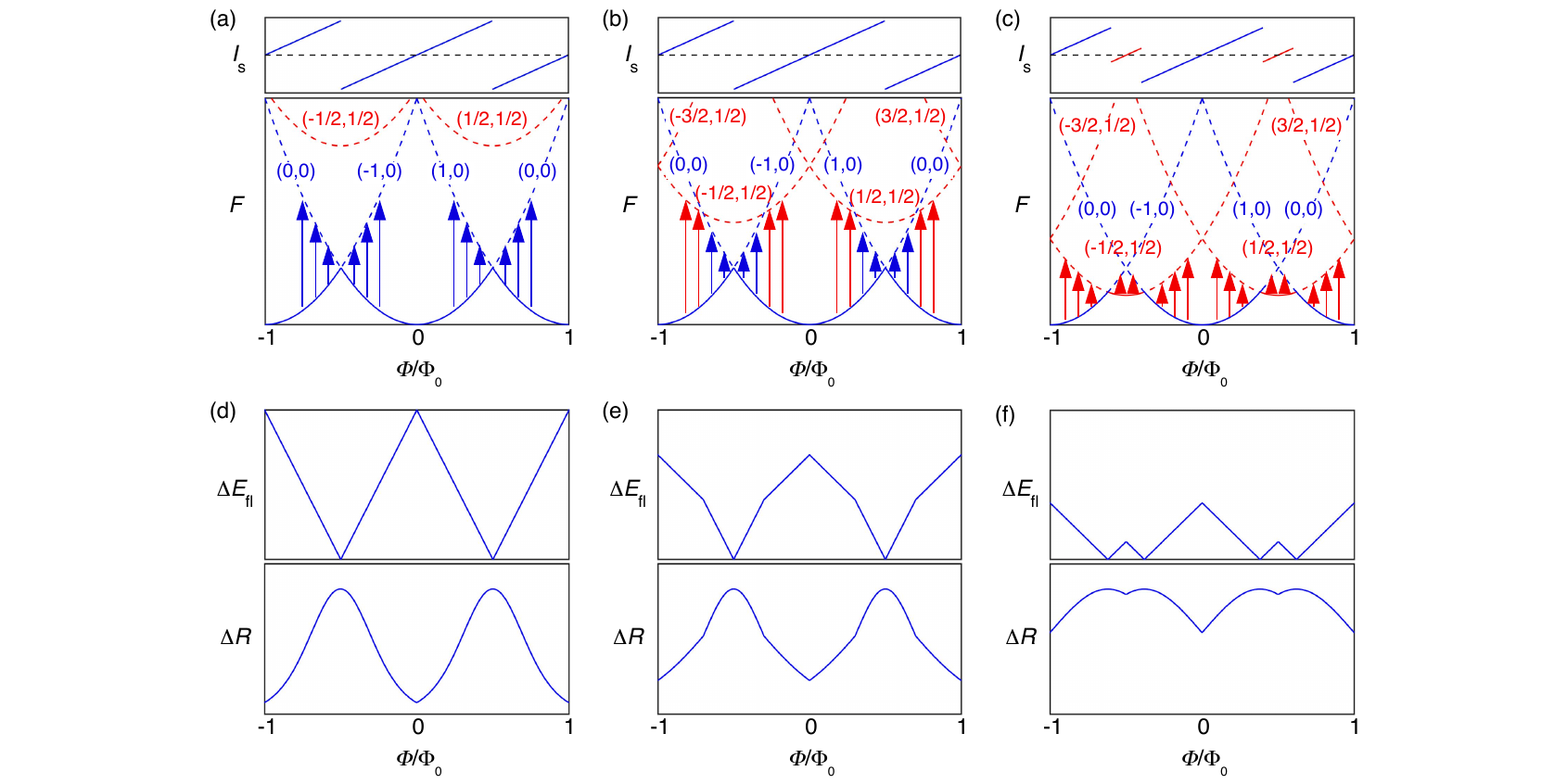} 
\caption{\label{Fig4} \textbf{$\Phi$ dependence of the free energy barrier for the vortex crossing and MR oscillations}. (a-c) Circulating supercurrent $I_s$ (Upper panels) and the kinetic energy part of the free energy $F$ (lower panel) of a doubly connected cylinder of a spin-triplet superconductor in the fluxoid state of $(n_s,n_{sp})$ in appropriate units based on Eq.~\ref{eq:freeenergy}. The dashed line in the upper panels indicate the zero value of $I_s$. Eq.~\ref{eq:freeenergy} was used to calculate $F$ as a function of $\Phi$. Three different combinations of $\rho_{sp}/\rho_{s}$ and $(1+\beta)^{-1}$ values (that determine respective free energies of HQV and FQV fluxoid states; see main text) were used. The measurement current $I_m$ was set to zero. When the free-energy parabolas for the HQV fluxoid state is lowered sufficiently by geometric constraints and applying a large $H_{||ab}$, the HQV states become the "ground" states near $\Phi = \Phi_0/2$ (red solid lines) as shown in (c). A transition from the "ground" to the "first excited" fluxoid state (dashed lines) at a fixed $\Phi$ is accompanied by the entry of an FQV or HQV into the interior of the cylinder, as indicated by arrows (blue for FQV and red for HQV transition). Returning from the "first excited" to the "ground" fluxoid state corresponds to the exit of an FQV or HQV with an opposite sign from the interior of the cylinder through the cylinder wall on the other side (arrows not shown). (d-f) (Upper panels) The free-energy difference between the "ground" and the "first excited" fluxoid state, $\Delta E_{fl}(\Phi)$, computed from the free energy differences shown in (a-c) (Lower panels). The free energy barrier for the vortex to enter and exit from the interior of the cylinder, $\Delta E_{in}(\Phi)$ and $\Delta E_{fl}(\Phi)$, differ from $\Delta E_{fl}(\Phi)$ by a constant. The expected MR oscillations, $\Delta R(\Phi)$, was calculated using Eq.~\ref{eq:AH} with $\Delta E=\Delta E_{fl}$. In (f), the crossings of HQVs is seen to lead to a dip feature in the MR peaks.
}
\end{figure*}


\subsection{Free energy barrier and MR oscillations at large $I_m$}


We argue below that the dip feature seen in the MR oscillations in Fig. 2 at large $I_m$ is a result of crossings of HQVs overcoming a free energy barrier which is a periodic function of $\Phi$, with the large $I_m$ playing the important role of tilting the barrier as illustrated in Fig.~\ref{Fig1}(e). For simplicity we will begin by considering the free-energy of a doubly connected cylinder featuring $w\ll r,\lambda_{ab}$, where $\lambda_{ab}$ is the in-plane penetration depth (with the magnetic field applied along the $c$ axis) with $I_m$ set to zero.
The kinetic-energy (or phase-gradient) part of the free energy of a doubly connected cylinder of a spin-triplet superconductor per length with $\Phi$ being varied is \cite{chung_stability_2007},
\begin{equation}
\label{eq:freeenergy}
F(n_s,n_{sp})=\left(\frac{\Phi_0^2}{8\pi^2 r^2}\right)\beta\left[\frac{1}{1+\beta}\left(n_s-\frac{\Phi}{\Phi_0}\right)^2+\frac{\rho_{sp}}{\rho_s}n_{sp}^2\right],
\end{equation}
where $\Phi = \pi r^2 H_{||c}$ is the applied magnetic flux enclosed in the cylinder, $\rho_s$ and $\rho_{sp}$ are the superfluid and spin fluid densities, and $\beta=rw/(2\lambda_{ab}^2)$.
It is seen from Eq.~\ref{eq:freeenergy} that the free energy difference of an HQV fluxoid state ($n_s$ = half integer) and that of FQV ($n_s$ = integer) depends on the relative values of $\rho_{sp}/\rho_{s}$ and $(1+\beta)^{-1}$, both of which are material properties dependent on the temperature. $(1+\beta)^{-1}$ also depends on the sample dimensions. A thin cylinder wall with a small $w$ will in general help lower the free energy of an HQV fluxoid state.
Note that the free energy shown in Ref.~\cite{chung_stability_2007} (Eq. 12) has an additional $(\Phi/\Phi_0)^2$ term, which will not make any difference in $\Delta E_{in}$, $\Delta E_{fl}$, or $\Delta E_{out}$ because only changes in the free energy between two different fluxoid states are relevant (see below).  

We will now show how the free energy are affected by the parameters of $\rho_{sp}/\rho_{s}$ and $(1+\beta)^{-1}$ by calculating the free energy of the cylinder using Eq.~\ref{eq:freeenergy} for three combinations of $\rho_{sp}/\rho_{s}$ and $(1+\beta)^{-1}$. We begin by plotting the circulating superfluid density as a function of applied $\Phi$ (Figs.~\ref{Fig4}(a)-(c) (upper panels))). It is seen that the sign of $I_s$ will reverse at at applied half-flux quanta \(\Phi = (m + \frac{1}{2})\Phi_0 \), where $m$ is an integer, or as the system enters/exits the HQV fluxoid state (Figs.~\ref{Fig4}(d), upper panels). The sign of the $c$-axis magnetic moment will change its sign accordingly, which enabled the detection of HQV fluxoid states in the original torque magnetometry experiment\cite{jang_observation_2011}.
For Fig.~4(a), $[\rho_{sp}/\rho_{s}]$/[$(1+\beta)^{-1}$]=3.08, corresponding to a sample in which the FQV fluxoid states (blue parabolas) are energetically favored over the HQV fluxoid states (red parabolas) at all values of $\Phi$. The free energy of the "ground" fluxoid state as a function of $\Phi$ is the same as that of a conventional $s$-wave superconductor \cite{tinkham_introduction_1996}. In Fig.~4(b), [$\rho_{sp}/\rho_{s}$]/[$(1+\beta)^{-1}$]=1.8, which lowers the free energy of the HQV fluxoid states (red parabolas) in comparison with that shown in Fig.~4(a). If we set [$\rho_{sp}/\rho_{s}$]/[$(1+\beta)^{-1}$ = 0.52, the free energy of the HQV fluxoid states will be lowered further 
(Fig.~\ref{Fig4}(c)). The same amount of lowering of the free energy can be achieved alternatively by applying an in-plane field with the same values of $\rho_{sp}/\rho_{s}$ and $(1+\beta)^{-1}$ used in Fig.~4(b). 
Fig.~4(c) shows that the presence of $H_{||ab}$, whose effects were not taken into account explicitly in Eq.~\ref{eq:freeenergy}, will push down the free energy of the HQV fluxoid states because of the coupling between the SSP accompanying the HQV fluxoid state and $H_{||ab}$\cite{vakaryuk_spin_2009,roberts_numerical_2013}. When $H_{||ab}$ is sufficiently large, the HQV fluxoid state would have a free energy lower than that of the FQV near \(\Phi = (m + \frac{1}{2})\Phi_0 \), where $m$ is an integer. In this case, the "ground states" would be either the HQV or the FQV fluxoid state as $\Phi$ is varied, as shown in Fig.~\ref{Fig4}(c). 

Excitations of the fluxoid states, from the ``ground" to the ``first excited" states, associated with the vortex crossing are indicated schematically by arrows in Figs.~\ref{Fig4}(a)-(c), in which the blue arrows mark the transitions with $n_s$ changing by 1, accomplished by the crossing of an FQV, and the red arrows mark those with $n_s$ changing by 1/2 induced by the crossing of an HQV. Note that the transitions from the ``first excited" back to the ``ground" fluxoid state are not shown in Figs.~\ref{Fig4}(a)-(c) to avoid the figures being too busy. In Figs.~\ref{Fig4}(d)-(f) (upper panels), the free energy spacing between the ``ground" and the ``first excited" fluxoid states, $\Delta E_{fl}$, are plotted as a function of $\it\Phi$, respectively. While the free energy of each fluxoid state itself is quadratic function of $\Phi$, interestingly, the ``excitaion energy" is a linear function of $\Phi$. The rate of the thermally activated vortex crossing will be essentially controlled by the free energy barriers for the vortex to enter and exit from the interior of the cylinder, $\Delta E_{in} (\Phi)$ and $\Delta E_{out} (\Phi)$. For the spin-triplet superconductors, the models for calculating $\Delta E_{in}(\Phi)$ and $\Delta E_{out}(\Phi)$ are not available. We expect that the effective free energy barrier is a periodic function of $\Phi$ with a form similar to $\Delta E_{fl} (\Phi)$ \cite{SM}. Here, to understand the periodic features in our MR data qualitatively, we calculate MR using Eq.~\ref{eq:AH} with the activation barrier determined by $\Delta E_{fl} (\Phi)$.

In the case of Fig.~\ref{Fig4}(d), the vortex crossing induced MR oscillations are conventional (lower panel).
In Fig.~\ref{Fig4}(e) (upper panel), the free energy   barrier for vortex crossing as a function of ${\it\Phi}$ has a kink at the point where the free energy parabola of the FQV and that of HQV fluxoid states coincide  (Fig.~\ref{Fig4}(b)). However, this subtle change of slope in MR is unlikely to be observed experimentally (Fig.~\ref{Fig4}(e), lower panel). Finally, in the presence of a sufficiently large $H_{||ab}$, the free energies of the HQV fluxoid states are substantially lowered as shown in Fig.~\ref{Fig4}(c). A peak in $\Delta E_{fl}$ found near ${\it\Phi}=(m+1/2)\Phi_0$ will lead to a dip in MR (Fig.~\ref{Fig4}(f)). 


The above analysis suggests that the dip feature seen in MR oscillations at large $I_m$ is a result of the HQV crossing (Figs.~\ref{Fig4}(f), lower panel), originating presumably from the corresponding feature in free energy barriers.

\subsection{MR oscillations at reduced $I_m$}

\begin{figure}[t]
\includegraphics{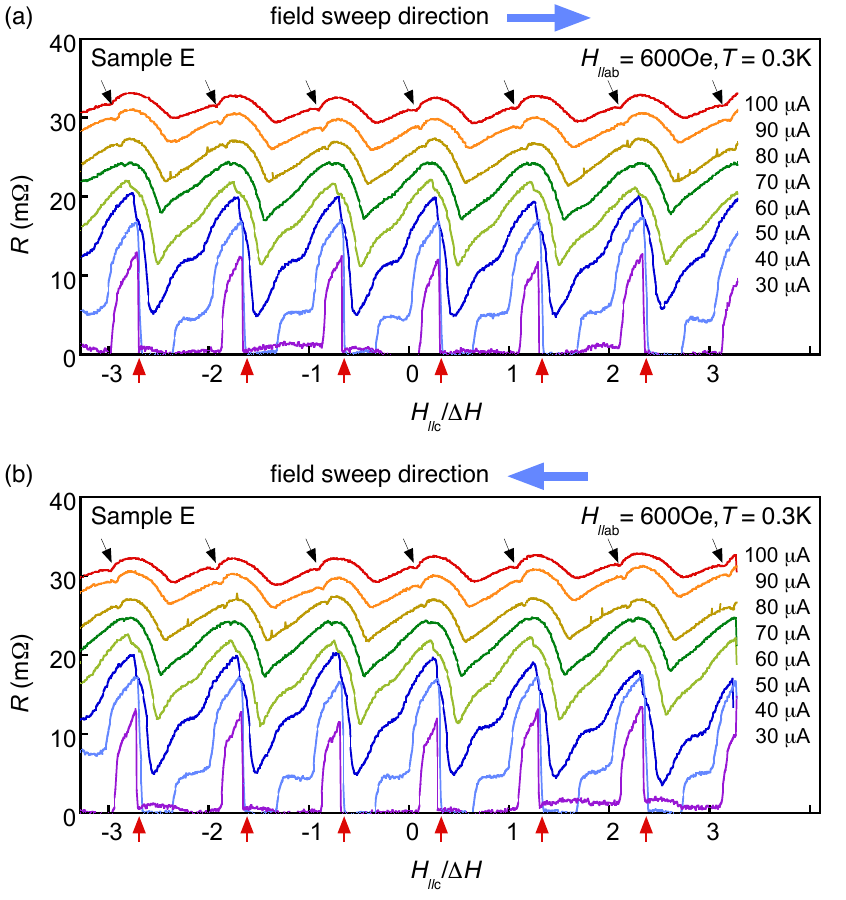}
\caption{\label{Fig5} \textbf{Anomalous features in MR oscillations at reduced $I_m$.} $R$~$vs$.~$H_{||c}$ for Sample E at a fixed in-plane field $H_{||ab}=$600~Oe for a wide range of applied current ($I_m$), as  $H_{||c}$ is swept from -123 to 123~Oe (a) and the other way around (b). Sharp drops in $R$, indicated by arrows, are clearly seen. The MR oscillation period is $\Delta H = 18.8$~Oe. A side peak in MR oscillations is also seen in a narrow range of $I_m$. Both the sharp sample resistance drop and the secondary MR peak are independent the direction of the $H_{||c}$ sweeping. 
}
\end{figure}

\begin{figure*}[t]
\includegraphics{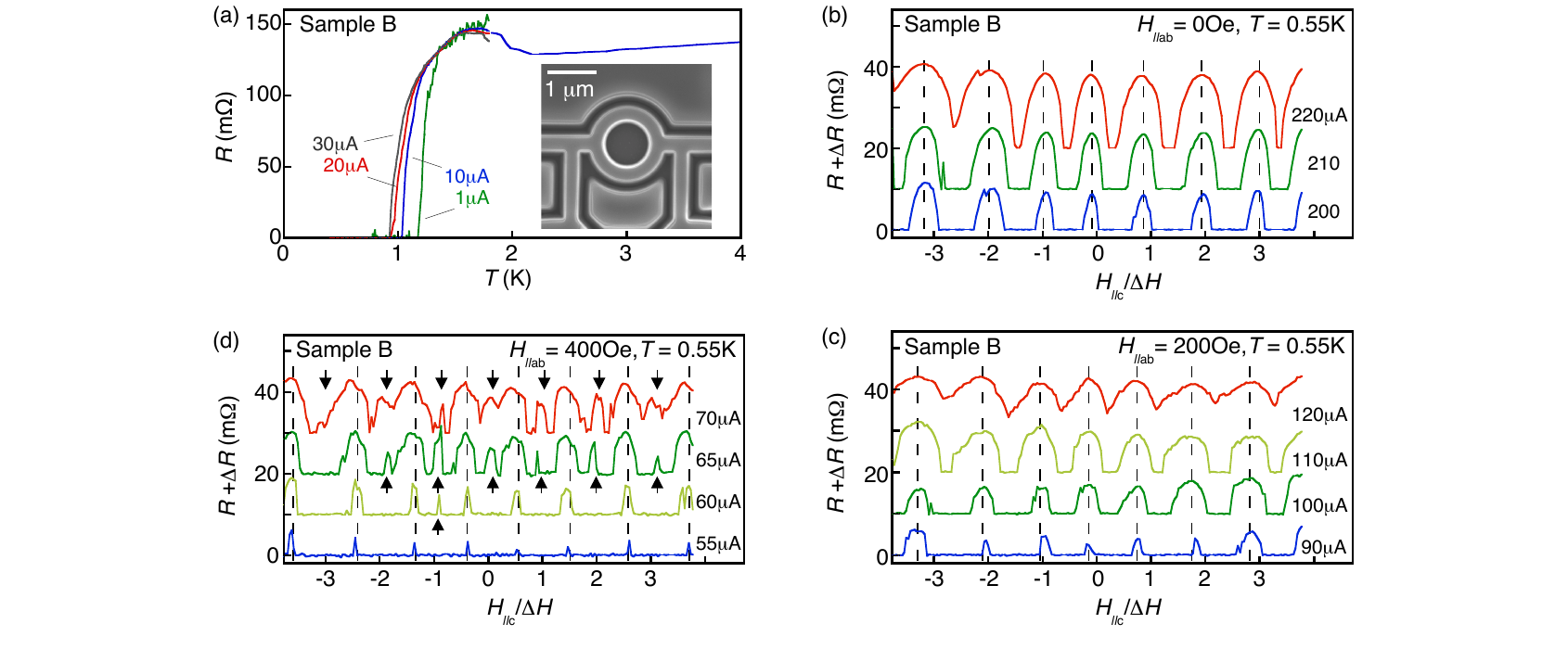}
\caption{\label{Fig6}\textbf{Secondary MR peaks at integer flux quanta}. (a) Sample resistance ($R$) $vs$.~temperature ($T$) for Sample~B, measured in zero magnetic field and at various $I_m$ values as indicated. Inset: An SEM image of the sample. $R$~$vs$.~$H_{||c}$ curves obtained at a fixed in-plane field of $H_{||ab} = 0$ (b), 200 (c) and 400~Oe (d) are shown with values of $I_m$ as indicated. $\Delta H = 16.3~$Oe (corresponding to $\Phi_0$). The curves are shifted vertically by $10~\Omega$ for clarity. The MR oscillations with a period of $\Phi_0$ is indicated by dashed lines. The secondary MR peaks found near the full flux quanta at $H_{||ab} = 400$~Oe (d) are indicated by arrows.  
}
\end{figure*}

\begin{figure}[t]
\includegraphics{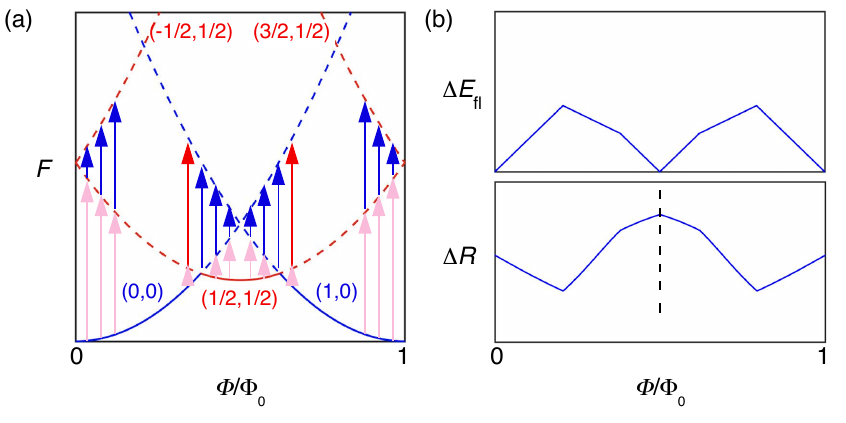}
\caption{\label{Fig7}\textbf{Possible origin of the secondary MR peak}. (a) $I_s$ (Upper panels) and $F$ (Lower panel) circulating current and free energy parabolas for fluxoid states of $(n_s,n_{sp})$ in a large in-plane field as shown in Fig.~\ref{Fig6}(c). Trapping of an HQV fluxoid state will allow crossings of an FQV along with transitions from the "first excited" to the "second excited" fluxoid states. The red arrows indicate the transitions accompanied by HQV crossings and the blue arrows those accompanied by FQV crossings. (b) Upper panel: The "excitation energy" as function of ${\it\Phi}$ for fluxoid state transitions shown in (a) taking into account of the temporary "trapping" of an HQV fluxoid state. Lower panel: Expected MR oscillations. The rates of the phase slips from the crossings of HQVs and FQVs were not distinguished and the MR was calculated using Eq.~\ref{eq:AH}. 
}
\end{figure}

As $I_m$ is lowered, the dip feature in Sample~E is seen to shift significantly (Fig.~\ref{Fig3} (a)). Reversing the direction of $I_m$ is seen to result in significant changes not only on the dip position but the amplitude of MR oscillations itself (Figs.~\ref{Fig3} (b-c)). In Figs.~\ref{Fig5}(a) and (b), we show the $R$~$vs$.~$H_{||c}$ curves for $H_{||ab} = 600$~Oe and fixed values of $I_m$. To ensure that no hysteresis from vortex trapping was involved, the $H_{||c}$ was swept from -123 to 123~Oe [Figs.~\ref{Fig5}(a)] and back [Figs.~\ref{Fig5}(b)], yielding essentially identical features. The dip position and its shift are seen to be independent of the direction in which $H_{||c}$ was swept, suggesting that the observed behavior is not due to flux trapping, which would normally be different when the field is swept in the opposite directions. As $I_m$ was lowered (from 120 to 60$\mu$A), the dip feature continues to shift towards the center of the MR peak while the position for the maximum of the MR peak remains essentially unchanged. 

More strikingly, as $I_m$ was lowered even further so that the dip feature approaches the maximum of the MR peak, it appears to turn into a sharp drop in the sample resistance as indicated by the arrows in Figs.~\ref{Fig5}(a) and (b). In addition, a side peak is seen to emerge on the side of the main MR peak where the dip feature was located when $I_m$ was larger. As shown in the SM\cite{SM} in more detail, this side peak becomes better defined as $I_m$ was lowered further even though its height continues to reduce. Meanwhile, the sharp drop in sample resistance becomes deeper and deeper until the resistance drops to zero. Interestingly, the sample resistance in the range of ${\it\Phi}$ in which the side peak in the MR is found also becomes zero at these smallest $I_m$.  

The corresponding values of $I_m$ at which the sample resistance started to drop sharply to zero, around 40-50~$\mu$A, are in fact not so small. Based on an analysis on a doubly connected $s$-wave superconductor with the same sample dimensions as Sample E (see SM\cite{SM}), these $I_m$ values are likely to be comparable with the maximal circulating current, $I_s^{max}$, expected near applied half-flux quanta. Therefore, these reduced $I_m$ values were in fact substantial.

It is clear that, as a function of ${\it\Phi}$, the free energy barrier characterized by $\Delta E_{in}$, $\Delta E_{fl}$, and $\Delta E_{out}$ possesses no sharp rises needed to produce a sharp drop in sample resistance as seen in Figs.~\ref{Fig2}(c) and (f) (upper panel). The decreasing tilting of the free energy barrier expected at reduced $I_m$ will not introduce a sharp rise in free energy either. On the other hand, the systematic evolution of the dip feature, the side peak in MR, and sharp drop in sample resistance seen in Sample E as $I_m$ was reduced seems to suggest that they share the same physical origin. 

Close inspection of Fig.~\ref{Fig2}(c) suggests that the "excitation" of the fluxoid state transitions from the "ground" to the "first excited" state lead to either an increase or a decrease in $n_s$. As shown in SM\cite{SM}, for a fluxoid state transition with an increase in $n_s$, the HQV that enters the cylinder has its sign of the magnetic flux the same as that of the applied magnetic flux, ${\it\Phi}$, which we refer to as a "positive" HQV. The returning of the cylinder from the "first excited" to the "ground" state will be associated with the exit of an HQV with the sign of the magnetic flux opposite to that of the applied magnetic flux, referred to as a "negative" HQV. For a fluxoid state transition with a decrease in $n_s$, similar analysis can also be carried out. The Lorentz force on a "positive" and "negative" HQV will point in opposite directions, causing the "potential barrier" as shown in Fig.~\ref{Fig1}(e) to tilt in opposite ways as well. As a result, the "positive" and the "negative" HQV will cross the sample in opposite directions. The sharp drop in sample resistance seen in Fig.~\ref{Fig5} can be understood if only a "positive" HQV or "negative" HQV can cross the sample at a reduced $I_m$, as described in more detail in SM\cite{SM}.

There could be many ways for the crossing of HQV of only a single sign. If $I_m$ is distributed asymmetrically between the two sides of the cylinder, the "positive" and the "negative" HQV entering the cylinder wall will be subjected to different $\Delta E_{in}$. The exit of the HQV with the sign opposite to the HQV entering the cylinder will encounter a $\Delta E_{out}$ larger than $\Delta E_{in}$ as well. As shown in SM\cite{SM}, a substantial asymmetrical distribution of $I_m$ will lead to a sharp drop in sample resistance at ${\it\Phi}$ values when the system enters or exits the region in which the HQV fluxoid is the ground state. However, in this scenario, a second sharp drop in sample resistance, smaller in magnitude compared with the major one, is also expected. Moreover, the sample resistance will not drop to zero. In addition, the side peak in MR cannot be explained by the asymmetrical distribution of $I_m$, as discussed in SM\cite{SM}.

Additional factors should also affect the crossing of HQV when $I_m$ is reduced and distributed asymmetrically between the two sides of the cylinder. For example, the HQV that has entered the cylinder can be trapped. The "positive" and the "negative" HQV can also be subject to different $\Delta E_{in}$. These additional factors may have resulted in the total "block" of the HQV crossing that led to the vanishing of the sample resistance for some ${\it\Phi}$. The vortex crossing in some ${\it\Phi}$ values be favored by a fortuitous combination of parameters. It follows that the range of $I_m$ in which the sharp drop to zero and the side peak in MR can be observed should be narrow, as seen experimentally. The side peak was seen roughly from 45 to 50~$\mu$A for $H_{||ab} = 400$~Oe [Fig.~\ref{Fig5}(c)] and 38 to 50~$\mu$A for $H_{||ab} = 600$~Oe [Fig.~\ref{Fig5}(d)] over a very narrow range of ${\it\Phi}$ values (see SM\cite{SM}). Interestingly, $R$=0 is seen in a wider range of ${\it\Phi}$ values at $H_{||ab} = 400$~Oe [Fig.~\ref{Fig5}(c)] than $H_{||ab} = 600$~Oe [Fig.~\ref{Fig5}(d)], which must be originating from the larger free energy barrier at those ${\it\Phi}$ values for $H_{||ab} = 400$~Oe. This is understandable as the ${\it\Phi}$-dependent $\Delta E_{in}$ at $H_{||ab} = 400$~Oe is in general larger than that at $H_{||ab} = 600$~Oe.

\subsection{Secondary MR peaks at integer flux quanta}


For a doubly connected cylinder of conventional $s$-wave superconductor, the magnitudes of $\Delta E_{in}$ and $\Delta E_{out}$ at $I_m$=0 depends on $\lambda_{ab}^2$ ($\sim \rho_s$), where $\lambda_{ab}$ is the in-plane penetration depth and $\rho_s$ is the superfluid density. The free energy potential barrier will be tilted differently on the two sides of the cylinder wall due to the asymmetrically distributed $I_m$. A suitable combination of parameters, $\Delta E_{in}$ and $\Delta E_{out}$, and the degrees of tilting by $I_m$, may force the system to be stuck, or "trapped" in the "excited" fluxoid state after the HQV enters the cylinder for a varying amount of time. Experimentally, even though the value of $\rho_s$ is not known, a large value of $J_c$ should be a good indication of a large $\rho_s$ based on results of $s$-wave superconductors\cite{tinkham_introduction_1996}. It so happens that value of $J_c$ for Samples E is large (Table~1), making "trapping" of an excited HQV fluxoid state likely.   

From Table~1 it is clear that Sample E features a large value of $J_c$. Certain features seen in the MR oscillations at low $I_m$ in Sample E hint the "trapping" of an HQV fluxoid state in this sample as described in SM\cite{SM}). We identified another sample, Sample~B, that also possessed a large $J_c$ value and a dimension similar to that of Sample~E (). In Fig.~\ref{Fig7}(a), $R$~$vs$.~$T$ for the sample measured in zero magnetic field and at various $I_m$ values was shown. It is seen that an anomalous resistance peak right above $T_c$ marked by the sharp drop in $R$. The existence of such a resistance anomaly, which is related to charge imbalance in mesoscopic superconductors, is well known and understood for conventional $s$-wave superconductors\cite{fink_critical_1985, WangLiu_Anomaly_2006}. Its physical origin has to do with an enhanced local $T_c$ at the voltage leads because of the multiple branching. In the present case, it is more likely that the enhanced local $T_c$ came from a combination of the existence of branching and dislocations as pointed above\cite{ying_enhanced_2013}. 

MR oscillations data measured on Sample~B at fixed in-plane fields are shown in Figs.~\ref{Fig7}(b-d). Asymmetry associated with $I_m$ also presented in this sample but not as strong as Sample~E\cite{SM}. The values of $I_m$ used for the MR measurements at $H_{||ab}=0$ are actually the highest among all samples used in this study, allowed by its very large $J_c$. At $H_{||ab} = 0$ and 200~Oe, smooth MR oscillations with a period of $\Phi_0$ were observed with no dip features. At $H_{||ab} = 400$~Oe, however, a set of secondary resistance peaks is clearly observed in addition to MR oscillations with a period of $\Phi_0$ [Fig.~\ref{Fig7}(d)]. Each secondary peak is seen to locate between two adjacent main MR peaks at applied integer flux quanta, ${\it\Phi}=m\Phi_0$, where $m$ is an integer.

The observation of secondary MR peaks near ${\it\Phi}=m\Phi_0$ at a large $H_{||ab}$ and $I_m$ is consistent with the picture in which an HQV fluxoid state was "trapped" in the sample temporarily that in turn enabled FQV crossings. In this scenario, when $H_{||ab}$ is sufficiently large, the free energy parabolas for HQV fluxoid states will be lowered substantially, making the "excitation energy" from the "ground" to the "first excited" fluxoid state substantially lower than that at $H_{||ab}$=0 even near ${\it\Phi}$ = $m\Phi_0$ (Fig.~\ref{Fig7}(a)). As a result, after the flux entered the cylinder to form the HQV fluxoid state ("first excited"), the system was trapped in this state temporarily. An FQV would then cross the cylinder wall and enable a transition from "first excited" to the "second excited" fluxoid state, followed by the exit of an FQV 
on the other side of the cylinder wall. Crossings of these FQVs would result in a secondary MR peak near ${\it\Phi}$ = $m\Phi_0$. Assuming that crossings of both "positive" and "negative" vortices are allowed in this sample, transitions from "first excited" to the "second excited" fluxoid state are all allowed, leading to the broadening of the main peaks in MR oscillations at ${\it\Phi}$ = $m\Phi_0$/2, as shown in Fig.~\ref{Fig7}(b). Importantly, transitions between two FQV fluxoid states near ${\it\Phi}=(1/2+m)\Phi_0$ enabled by the crossings of FQVs will feature no dip features in the main MR peaks, as seen experimentally.



\section{Discussion}

\subsection{Experimental considerations}

In the original torque magnetometry experiment on doubly connected, mesoscopic single crystals of Sr$_2$RuO$_4$, only the "ground" fluxoid states were explored through the measurement of the $c$-axis magnetic moment of the sample that is proportional to the total circulating current, $I_s$, at the given applied flux, $\it\Phi$ \cite{jang_observation_2011}. The experiment had the distinct advantage of being very sensitive. In addition, the sample temperature could be raised easily using a laser that changed the temperature of the sample only and cooled it down again with the cryostat, making it convenient to always field cool the sample for any $H_{||c}$ and $H_{||ab}$. In the current work, the sample was cooled at a fixed $H_{||ab}$ (including zero) with $H_{||c}$ swept at a fixed low temperature to avoid the very long time needed to warm our dilution refrigerator to above the $T_c$ of the sample and coll it down again to a low temperature. As a result, our sample may not always be at the state featuring the lowest free energy, possibly making our MR oscillations at times out of "sync" with $\Phi$ dependence of the free energy barrier, which may explain why an in-plane field much larger than those used in magnetometry measurements were needed in our MR oscillations measurements to reveal features associated with the HQV fluxoid state. It may also account for the observation that the dip feature was missing sometimes. In short, a large in-plane field can help enlarge those HQV features in the free energy barrier and make them observable in the MR oscillations.  

We have shown in this paper that the MR oscillations observed in the present work are due to measurement current driven vortex crossing rather than the LP effect. It was reported previously\cite{yasui_little-parks_2017} that MR oscillations observed in two square loops of single crystalline Sr$_2$RuO$_4$, which were prepared also by FIB, in the presence of an in-plane magnetic field were due to the LP effect. As discussed in more detail in SM\cite{SM}, the MR data presented in this previous work, which unfortunately contained excessive irregularities, is also consistent with the same picture presented above, including, most importantly, the existence of the HQV fluxoid state (see SM\cite{SM}). 
The presence of an in-plane magnetic field necessary for the observation of the HQV brings certain complications. In particular, the formation of in-plane vortices and their motion do have consequences. In the FQV splitting scenario of HQV formation in Sr$_2$RuO$_4$ in which the two Abrikosov HQVs are linked by a d-vector soliton\cite{kee_half-quantum_2000}, both $c$-axis and in-plane Abrikosov HQVs are in principle possible. However, in the present sample geometry, in-plane Abrikosov HQVs may be less likely to form than $c$-axis HQVs due to the lower degree of geometric constraints.  The free-energy barrier for an in-plane HQV/FQV to cross the doubly connected part of the sample should still be a periodic function of the $c$-axis applied flux. This free energy barrier will be tilted by the measurement current as well. Crossings of an in-plane HQV/FQV will generate a finite voltage across the sample, adding to the voltage produced by crossings of $c$-axis vortices. 

The formation of an in-plane vortex between the two voltage leads and the cylinder itself may cost additional free energy as it is known that the presence of multiple branching points in a mesoscopic superconducting sample tend to feature an enhanced superconducting order parameter, as shown previously in an $s$-wave superconducting structures\cite{fink_critical_1985}. In the current experiment, the enhancement of the superconducting energy gap in singly connected parts of the sample would make it less likely for a vortex to form and cross through them. In short, the contribution to the MR oscillations from the crossing of in-plane vortices is small.

\subsection{Stability of the HQV}

In the absence of an in-plane magnetic field, the free energy difference between an HQV fluxoid state and that of an FQV at applied half flux quanta is proportional to $\rho_{sp}/\rho_{s}$-$(1+\beta)^{-1}$. Only when $\beta$ $<$ $\rho_{s}/\rho_{sp}$-1 is an HQV fluxoid state stable near applied half flux quanta. Given that $\beta$ itself is proportional to $\rho_s$, whether a large $\rho_s$ in similarly sized samples will help the stability of the HQV depends on how $\rho_{sp}$ and $\rho_{s}$ are related to one another. Their relationship was analyzed in superfluid $^3$He\cite{leggett_theoretical_1975}. However, it is not well understood for spin-triplet superconductors. 

Among the four samples used in the present work, Samples E and B possess a large $J_c$ value. For these two samples, features associated with the HQV were clearly seen at $H_{||ab}$=400~Oe. Unfortunately, for a spin-triplet superconductor, how $\rho_{s}$ is related to $J_c$, which has been rarely studied theoretically\cite{KeeMaki_Ic}, is not understood. Experimentally, it is difficult to perform measurements on $\rho_s$ for a mesoscopic sample. In any case, if a large $\rho_{s}$ indeed favors the formation of HQVs, $\rho_{sp}$ and $\rho_{s}$ must be relatively independent, at least in our mesoscopic samples.     

Given that the presence of an in-plane field is by far the most important factor for stabilizing the HQV fluxoid state in mesoscopic Sr$_2$RuO$_4$, it is clear that the free energy of the HQV fluxoid state is lowered by coupling the spontaneous spin polarization to the in-plane magnetic field, which also implies that the spontaneous spin polarization associated with the HQV fluxoid state must have an in-plane component. In the original Vakaryuk-Leggett picture of ESP formation of HQV\cite{vakaryuk_spin_2009}, the d-vector was assumed to be in the in-plane direction, which means that the spontaneous spin polarization will be along the $c$ axis given that the formation of the HQV in the ESP is built on the mismatch in vorticity between the spin-up and spin-down species of the Cooper pair. Since an in-plane magnetic field does not couple to spontaneous spin polarization along the $c$ axis, the observed enhancement of the stability of HQV by an in-plane magnetic field in mesoscopic Sr$_2$Ru$O_4$ suggests that the d-vector in our mesoscopic samples must feature a substantial $c$-axis component. Among all pairing states allowed by the crystalline symmetry of Sr$_2$Ru$O_4$ in the Rice-Sigrist scheme\citep{rice_sr_1995}, only the $\Gamma_5^-$, a spin-triplet, chiral $p$-wave state, features a d-vector along the $c$ axis. However, it should be cautioned that the observed $c$-axis component of the d-vector can also originate from the d-vector rotation near the sample surface (as opposed to a $\Gamma_5^-$ pairing symmetry), where spin-orbital interaction and the crystalline symmetry are both different from those in the bulk.    

An energy efficient way of creating an HQV fluxoid state is to create a d-soliton in the cylinder wall and trap (2n+1)$\Phi_0/2$, where n is an integer, in the interior of the cylinder at the same time\cite{kee_releasing_2013}. The formation of an HQV through such a d-soliton avoids the cost in the condensation energy due to the normal core as an Abrikosov HQV would have to and saves the kinetic energy by relaxing geometrical constraints on the circulating charge supercurrents. The Ginzburg-Landau theory of the d-soliton predicts a spontaneous spin polarization with components along both the $c$ axis as well as the in-plane directions because of d-vector texture formed in the sample\cite{kee_releasing_2013}. Taking into account the Fermi liquid corrections, the Ginzburg-Landau theory was formulated to include only the radial component of the spontaneous spin polarization, $s_r$, in the theory by a specific choice of the gauge. Note that $s_r$ is nonzero only near the d-soliton as opposed to being distributed uniformly in the Vakaryuk-Leggett picture\cite{vakaryuk_spin_2009}. The in-plane magnetic field will then couple to the in-plane component of the spin polarization, thus stabilizing the HQV fluxoid state\cite{kee_releasing_2013}. As to be discussed below, $I_m$ may also couple to $s_r$.

\subsection{Spontaneous spin polarization (SSP)}

It is clear that the SSP plays a crucial role in stabilizing the HQV fluxoid state. In the Ginzburg-Landau theory of the d-soliton formation of the HQV fluxoid state, introduction of the phenomenological parameter $s_r$ leads to a shift in the free energy parabolas of the HQV fluxoid state. The parabolas now reach the minimum at ${\it\Phi}=\Phi_m$\cite{kee_releasing_2013}, deviating from ${\it\Phi}=m{\Phi_0}/2$ by an amount proportional to the square root of the strength of the spin-orbit coupling and $s_r$\cite{kee_releasing_2013}. This deviation in turn leads to a shift in the dip feature from ${\it\Phi}=m{\Phi_0}/2$. As shown in Fig.~\ref{Fig2} (g) and (h), the dip feature in the MR oscillations at a fixed $I_m$ in Sample E is seen to locate on the same side of the MR peak, independent of the sign of $H_{||c}$. This is easily understood within this Ginzburg-Landau theory: Since $\Phi_m$ is independent of $\it\Phi$, the resulted shift in the dip feature at a fixed $I_m$ is also independent of the sign of $\it\Phi$. 

In Fig.~\ref{Fig3}(a), the position of the dip feature, ${\it\Phi}=\Phi_m$, is shifted away from the maximal value of the main MR peak at ${\it\Phi}=m{\Phi_0}/2$ as $I_m$ was increased. Given that the spin-orbital coupling is unlikely to be affected by $I_m$, this shift in the dip feature must have originated from a change in $s_r$, whose magnitude at a fixed $H_{||ab}$ appears to increase as $I_m$ increases. As $s_r$ is only non-zero near the d-soliton, its effect on shifting the entire free energy parabolas of the HQV fluxoid state is somewhat unexpected. On the other hand, the switching of the dip feature from one side of the main MR peak to the other when the direction of $I_m$ was reversed, a trend seen in both Samples BL and E (Figs.~\ref{Fig3}(b) and (c)), is fully consistent with the picture - the shift of the dip feature with the varying $I_m$ is originated from changes in $s_r$, which is in turn affected by $I_m$.


A nonuniform distribution of the circulating component of the asymmetric measurement current, $I_m,cir$, along the axis of the cylinder (the vertical direction) may affect $s_r$. In the London limit, supercurrents flowing through a superconductor tend to concentrate near the sample surface. Because of the uneven thickness of the cylinder wall (thinner on the top and thicker at the bottom because of the way the cylinder was cut, as shown in SM\cite{SM}), we suspect that $I_m,cir$ will flow through the cylinder with the current density concentrating at the bottom surface of the sample. According to the Biot-Savart law, the direction of the magnetic field generated by a current is determined by how the current is distributed. For example, it is known that for an infinitely long, uniform solenoid, the magnetic field will be along the axis of the cylinder with zero radial component. In a infinitely thin loop, however, a large radial component of magnetic field is expected away from the current itself. We speculate that the effect of $I_m$ on $s_r$ is due to the asymmetrical, nonuniform distribution of $I_m$ in the sample. If $I_{m,cir}$ is concentrated at the bottom surface of the cylinder, it would have generated a radial component of the magnetic field in the rest of the sample according to the Biot-Savart law. As a result, $s_r$ will follow $I_{m,cir}$: when $I_m$ reverses its direction, the direction of the radial component of the field, and therefore, $s_r$, will be reversed; when $I_m$ is increased in magnitude, $s_r$ will increase as well. Taking into account all this, the behavior of $\Phi_m$ seen experimentally ma be understood.

\subsection{Pairing symmetry in Sr$_2$RuO$_4$}

A spontaneous half magnetic flux quantum of $\Phi_0/2$, the same as that in the HQV fluxoid state, is found in a very different context, with a  physical origin different from than discussed above. Proposed originally by Geshkenbein, Larkin, and Barone (GLB) \cite{GLB_half-h/2e-Theory_1987}, a hybrid superconducting quantum interference device (SQUID) consisting of two oppositely faced (oriented 180-degree apart) Josephson junctions between a spin-singlet $s$- and spin-triplet $p$-wave superconductor can be used to test the $p$-wave pairing symmetry. Here the two Josephson junctions feature Josephson couplings with opposite signs, corresponding to an intrinsic phase shift of $\pi$ across one of the two Josephson junctions in the loop. As a result, the phase winding accumulated by the phase gradient in the loop away from the two junctions will be $\pi$ instead of 2$\pi$, which leads to a spontaneous magnetic flux of $\Phi_0/2$. Experimentally, a structure of the GLB type was first realized in spin-singlet, $d$-wave superconductor of a high-$T_c$ cuprate in the form of a tricrystal loop or thin film \cite{tsuei_pairing_2000,schneider_half-h/2e_2004} as well as a hybrid corner SQUID or junction involving an $s$-wave and $d$-wave cuprate. In the latter samples, a phase shift of $\pi$ (or a flux shift of $\Phi_0/2$) in the quantum oscillations which results in a minimum critical current or maximum sample resistance at zero enclosed flux was found \cite{van_harlingen_phase-sensitive_1995}. Similar features appear to have also been observed in a hybrid Nb-granular iron pnictide superconducting loop \cite{ChenTsuei_half-h/2e_2010} as well as a ring of granular $\beta$-Bi$_2$Pd \cite{LiChien_half-h/2e_2019}. 

The physical origin of this spontaneous half magnetic flux seen in GLB structures is the "anomalous" Josephson effect between two unconventional superconductor or an unconventional and a conventional superconductor, which is very different from the HQV fluxoid state found in a doubly connected single crystalline, spin-triplet superconductor whose physical origin is the splitting of the phase winding between spin and orbital parts of the Cooper wave function, or the counterflow of the "spin-up" and "spin-down" species of the Cooper pairs in the ESP picture. The observation of the HQV in mesoscopic Sr$_2$RuO$_4$ provides robust, unambiguous evidence that Sr$_2$RuO$_4$ is a spin-triplet superconductor. Given that the crystalline symmetry of Sr$_2$RuO$_4$ is D$_{4h}$ featuring an inversion symmetry, the symmetry of the orbital part of the superconducting order parameter must be of the odd-parity, independent of the strength of the spin-orbital coupling and the presence of multiband Fermi surface. The latter of which assumes that the interband scattering is sufficiently large to exclude different pairing symmetry on different bands\cite{LeggettLiu_2020}. Odd-parity superconductivity has already been demonstrated in Sr$_2$RuO$_4$ previously by several authors of the current paper and their collaborators using phase-sensitive measurements\cite{nelson_odd-parity_2004}. As pointed out earlier in the paper, the necessary use of the in-plane magnetic field in stabilizing the HQV fluxoid state appears to favor a $c$-axis oriented d-vector. Even though only $\Gamma_5^-$ among the symmetry allowed odd-parity, spin-triplet pairing state features a d-vector along the $c$ axis\cite{Mineev_1999}, this current and previous HQV experiments by themselves do not necessarily exclude any of the helical state featuring an in-plane d-vector. The d-vector seen in the mesoscopic samples of Sr$_2$RuO$_4$ could feature a d-vector texture, in particular, near the boundary of the sample\cite{LeggettLiu_2020}.    

We note that the pairing symmetry of Sr$_2$RuO$_4$ is not settled\cite{MackenzieMaeno_2003,Annett_2004,Maeno_2012,Kallin_2012,Liu_2015,LeggettLiu_2020}. For many years since the publication of an intriguing experiment of muon spin rotation ($\mu$SR)\cite{LukeUemura_1998} and the original Knight shift measurements\cite{ishida_spin-triplet_1998} on Sr$_2$RuO$_4$ in 1998, the prevailing view in the community is that the $\Gamma_5^-$, a time-reversal-symmetry-breaking (TRSB) state with a d-vector perpendicular to the in-plane direction is the preferred choice for the pairing symmetry in Sr$_2$RuO$_4$. The $\Gamma_5^-$ state was also favored by the result of the selection rule in Josephson effect published in 2000\cite{JinLiu_2000} and the polarized neutron measurements\cite{DuffyHayden_2000}. Importantly, the $\mu$SR experiment was confirmed in 2012\cite{ShirokaForgan_2012}. The existence of a superconducting TRSB state in SRO received a further boost from the Kerr rotation experiment\cite{XiaKapitulnik_2006}, which supports the existence of a TRSB state in Sr$_2$RuO$_4$ even though issues related to the $\mu$SR and Kerr rotation experiments do exist, which are yet to be resolved\cite{LeggettLiu_2020}. On the other hand, the phase-sensitive experiments of some of the present authors and collaborators\cite{nelson_odd-parity_2004} can be interpreted only in an odd-parity, spin-triplet pairing picture. 


New Knight shift measurements were carried out on Sr$_2$RuO$_4$ with an in-pane magnetic field at the pulse energy smaller than that used previously\cite{pustogow_constraints_2019,Ishida_2019}, which revealed a distinct reduction in the Knight shift below T$_c$. It turns out that the previous finding\cite{ishida_spin-triplet_1998} showing that the Knight shift in Sr$_2$RuO$_4$ was a constant across T$_c$ and all the way down to the lowest temperature was in fact due to sample heating instead of the intrinsic properties of Sr$_2$RuO$_4$. The Knight shift was found to drop below T$_c$, reaching a value of (30 - 40)$\%$ of the normal-state value. This result was further supported by the new polarized neutron scattering experiment\cite{PetschHayden_2020}. These new findings raised doubts as to whether bulk Sr$_2$RuO$_4$ indeed possesses a pairing symmetry characterized by Sr$_2$RuO$_4$, or even is spin-triplet to begin with. Nevertheless, these new results do not rule out spin-triplet pairing even though they do exclude the $\Gamma_5^-$ with a $c$-axis d-vector within the spin-triplet scenario. It should be noted, however, that the spin susceptibility is defined in the zero magnetic field limit. Both the Knight shift and polarized neutron scattering measurements were carried out in a finite magnetic field. Therefore, strictly speaking, the the spin susceptibility has not been measured directly. Only when no second phase transition as the magnetic field is increased at low temperatures are the Knight shift and polarized neutron scattering results related directly to the spin susceptibility defined in the limit of zero magnetic fields\cite{LeggettLiu_2020}. Both HQV and the phase sensitive experiments are carried out near zero magnetic fields while the Knight shift and polarized neutron scattering measurements were done at a substantially large magnetic field. if the d-vector is along the $c$ axis in the surface region due to d-vector rotation, or alternatively, the $\Gamma_5^-$ state favored by the $\mu$SR and Kerr rotation experiments is realized near zero magnetic field but a helical state prevails in high magnetic fields\cite{LeggettLiu_2020}, these seemingly contradicting experimental results on Sr$_2$RuO$_4$ that are discussed above may be reconciled.

We would like to note that an HQV fluxoid state can also be formed from chiral domains and domain walls expected in a chiral $p$-wave, $\Gamma_5^-$ state\cite{kee_releasing_2013}. It was predicted\cite{kee_releasing_2013,XiaoHu_2015} that the formation of the domain walls in the cylinder will lead to the formation of a fractional quantum vortex (including HQV) in general. Experimentally, the formation of chiral domains in Sr$_2$Ru$O_4$ was explored previously, $e.g.$, in a Josephson junction study of time-reversal symmetry breaking in Sr$_2$RuO$_4$\cite{kidwingira_dynamical_2006}. However, data obtained in the present work seems to suggest that only HQV and FQV were formed.  


\subsection{Abrikosov HQVs}

The HQV crossing picture described above requires that the magnetic flux penetrates into the cylinder wall for a certain period of time. The penetrated flux turns into an Abrikosov HQV in the cylinder wall. For the d-soliton picture described above, a d-soliton is formed at the same time. Similar to an Abrikosov FQV, an Abrikosov HQV is envisioned to be a singularity of vanishing order parameter, which turns into a "normal core" at finite temperatures. The size of the vortex should be on the order of the in-plane penetration depth and the size of the normal core is that of the superconducting coherence length. Both characteristic lengths are temperature dependent, diverging approaching to T$_c$. In addition, similar to FQV, the "normal core" of an Abrikosov HQV will also feature discrete core states. It was proposed that an Abrikosov HQV hosts a Majorana zero energy mode (MZEM) which is useful for fault tolerant topological quantum computing (TQC)\cite{das_sarma_proposal_2006} . 

An Abrikosov HQV may be moved around in the sample by a current induced Lorentz force. The ability to move an HQV could be important for implementing TQC which relies on the braiding operations. At a finite temperature, the motion of an Abrikosov HQV is subject to a damping force originating from the exchange of momentum between the core states and the quasiparticles in the superconductor, similar to an FQV in a conventional superconductor\cite{tinkham_introduction_1996}. The damping force of the Abrikosov HQV should be different from that of the Abrikosov FQV in a conventional superconductor because of the different energy spectrum for the vortex core states. The study of the damping force on the Abrikosov HQV will require samples with a geometry different from that used in the present study.



\section{Conclusions}

To conclude, our MR oscillations measurements on doubly connected cylinders of single crystalline Sr$_2$RuO$_4$ of a mesoscopic size carried out at a large in-plane magnetic field revealed distinct features that supports the existence of the HQV. This work provides additional evidence for odd-parity, spin-triplet pairing in Sr$_2$RuO$_4$, favoring a $c$-axis oriented d-vector. This work also provides insights into the stability and dynamics of the HQV. The confirmation of the existence of HQVs in mesoscopic Sr$_2$RuO$_4$ revealed an additional analog between Sr$_2$RuO$_4$ and superfluid $^3$He given the recent success in observing HQVs in the latter in which geometrical constraints appeared to also play an important role\cite{autti_observation_2016}. Important questions to be investigated in the future include the quantitative determination of the SSP and the associated spin fluid density, the direct imaging of an HQV, the demonstration of the non-Abelian statistics of MZEMs bound to an HQV as well as the measurement of the damping force on an HQV that are important for the potential use of HQVs for fault tolerant TQC.

\begin{acknowledgments}
We would like to thank A. J. Leggett, V. Vakaryuk, S.-K. Chung, K. Roberts, Y. Maeno, J. Kirtley, C. Kallin, J. A. Sauls, J. K. Jain, Z. Wang and W. Huang for useful discussions. The work done at Penn State was supported by DOE under Grant No. DE-FG02-04ER46159. Z.Q.M. acknowledges the support by the US National Science Foundation under grant DMR 1917579. 
  
\end{acknowledgments}


\bibliography{HQV}

\end{document}